\documentclass[journal]{vgtc}                     % final (journal style)

\usepackage{graphicx} % Required for inserting images
\usepackage{tabularx}
\usepackage{array}
\usepackage{longtable}
\usepackage{lipsum}
\usepackage{gensymb}
\usepackage{hxlab-macros} % QoL macros reside here
\usepackage{sparkline-macros}
\usepackage{subcaption}
\usepackage{wasysym}
\usepackage[table]{xcolor}

\definecolor{lightgray}{gray}{0.9}

\onlineid{1590}

\vgtccategory{Research}

\vgtcpapertype{Area 1: Theoretical \& Empirical}

\title{Vibes on Demand: Adding Vibrotactile Encoding to Line Charts Shows Experiential Benefits Without Performance Costs}

\author{%
  Anchit Mishra, Oliver Schneider, and Matthew Brehmer
}
\sethlcolor{red!25}
\renewcommand\hl[1]{#1}
\renewcommand\sout[1]{}
\renewcommand\st[1]{}
\authorfooter{
  \item
  	Anchit Mishra, Oliver Schneider, and Matthew Brehmer are with the University of Waterloo.
  	E-mail: \{amishra,oliver.schneider,mbrehmer\}@uwaterloo.ca
}

\abstract{%
Details on demand is a common design pattern in visualization design, especially useful when interacting with visually-saturated or small displays. Beyond visualization, another common approach for saturated displays is to incorporate other modalities, such as haptic feedback. While haptic rendering in visualization has primarily targeted accessibility needs, with haptics as a \textit{substitute} for visual feedback, studies using haptics outside of a visualization context have shown value in experiential factors, such as increased confidence in ambiguous contexts and higher engagement. We explore vibrotactile feedback as a \textit{reinforcing} information channel for communicating trends in details-on-demand tooltips on touchscreens. We identify preferred parameter configurations for our haptic encoding, informed by a study where participants identified parameter configurations that they perceived to most accurately reflect the dynamics of line charts appearing in tooltips. In a second study, we evaluated participant performance in a pairwise comparison task, finding that incorporating vibrotactile encoding improves involvement without affecting accuracy. We discuss the implications of these findings for future visualization design, and propose directions for applications and future studies.
  }

\keywords{Haptics, multi-sensory interfaces, details-on-demand, temporal data, perception \& cognition, human-subjects studies.}

\teaser{
  \centering
  \includegraphics[trim={0 3.2cm 0 0}, clip, width=\textwidth]{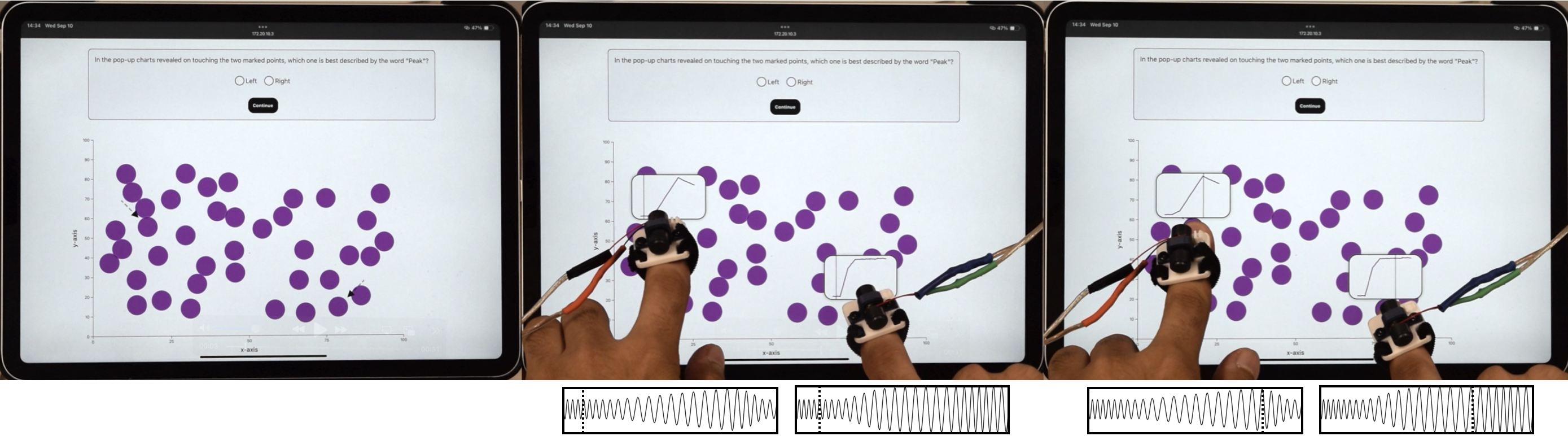}
  \caption{We study details-on-demand visualization interfaces augmented with vibrotactile feedback for simultaneous value comparison. When tapping on a mark in the scatterplot, its corresponding line chart appears and its vibrotactile counterpart is felt repeatedly via a finger-mounted actuator while a visual playhead traces the playback. With two fingers, vibrotactile playback is 
  synchronized in phase. }
  \label{fig:teaser}
}

\graphicspath{{figs/}{figures/}{pictures/}{images/}{./}} % where to search for the images

\usepackage{tabu}                      % only used for the table example
\usepackage{booktabs}                  % only used for the table example
\usepackage{lipsum}                    % used to generate placeholder text
\usepackage{mwe}                       % used to generate placeholder figures

\usepackage{mathptmx}                  % use matching math font

\begin{document}

%%%%%%%%%%%%%%%%%%%%%%%%%%%%%%%%%%%%%%%%%%%%%%%%%%%%%%%%%%%%%%%%
%%%%%%%%%%%%%%%%%%%%%% START OF THE PAPER %%%%%%%%%%%%%%%%%%%%%%
%%%%%%%%%%%%%%%%%%%%%%%%%%%%%%%%%%%%%%%%%%%%%%%%%%%%%%%%%%%%%%%%

%% \firstsection{Introduction}

\maketitle

%% \section{Introduction} %for journal use above \firstsection{..} instead
\section{Introduction}

Visualization interfaces often incorporate a \textit{details-on-demand} interaction design pattern~\cite{shneiderman2003eyes}, such as by revealing a modal window or tooltip when selecting a visual data element in a chart or dashboard~\cite{bach2022dashboard}. 
In practice, details-on-demand views \st{will in turn}present visual representations of selected elements~\cite{deng2023revisiting}.
These views are often by necessity small\st{, so as} to preserve the context of the underlying chart, and are thus populated with word-scale \cite{goffin2014exploring} micro-visualizations \cite{isenberg_habilitation} \st{such as}\hl{like} sparklines \cite{tufte_beautiful_evidence}. 
Despite the prevalence and commercial tool support for this design pattern (\eg~in Tableau~\cite{tableau_data-viz_2017} and Microsoft Power BI~\cite{powerbi2024}), the analysis and decision-making experience of these compact and abstract visual representations in small or visually-saturated displays has received little research attention.

One unexplored opportunity with respect to supporting details-on-demand on small or visually-saturated displays lies in the haptic feedback that these devices can provide.
Prior work in perceptual psychology shows that the sense of touch can impart additional benefits, such as by eliciting greater confidence in one's decisions when faced with visual ambiguity \cite{fairhurst2018confidence}, by enabling faster task performance \cite{corbett2016effects}, or by improving accuracy \cite{kreimeier2019evaluation}.
However, prior visualization applications incorporating haptic feedback typically serve to assist blind and low-vision (BLV) individuals as they make sense of data that would otherwise be represented visually, such as by using shape-changing displays to render surface-projected visuals~\cite{follmer_inform_2013} or by generating tactile glyphs~\cite{roberts_haptic_2005} to communicate multiple attributes of data elements.

In this paper, we propose the use of vibrotactile haptic feedback in details-on-demand views as a 
\textit{reinforcing} visualization design pattern for small visually-saturated touchscreen displays, rather than as a \textit{substitutive} visualization design pattern.  
We explore this approach with a focus on glanceable line charts representing continuous time-oriented data.
We describe the generation of \textit{vibrotactile line charts} revealed via details-on-demand interaction\hl{, as shown in}~\autoref{fig:teaser}, and then investigate their characteristics in two studies.
The first ($n = 12$) probes the design space of vibrotactile parameters for representing time-series data, in which we asked participants to tune parameters until they arrived at a configuration that they determined to best match a corresponding visual chart. 
From the study results, we derived a vibrotactile encoding parametrization. Given the compact and glanceable nature of our line charts as opposed to full-fledged line charts, we investigate vibrotactile feedback as a means to saliently convey temporal trends and features such as peaks and troughs, rather than as a means to allow for precise value retrieval.
In a second study ($n = 21$), we presented participants with a pairwise comparison task \st{so that we might}\hl{to} better understand the effectiveness of vibrotactile feedback \st{as a means to compare}\hl{for comparing} temporal features shown in details-on-demand line charts.
Here, we asked participants to select the line chart that best matched a given feature descriptor term.
We found that the addition of vibrotactile feedback does not reduce accuracy, and that it adds experiential value, suggesting a potential to direct attention in ways that support deliberation and decision-making grounded in data. 
Finally, we discuss 
the implications of our findings for future visualization design.

\section{Related Work}

Our work is informed by the \textit{details-on-demand} visualization design pattern and by prior work in visualization beyond conventional desktop displays. It is also informed by research in haptic perception and technology, which we explored as an additional modality for visualizing data on small or visually-saturated displays.

\subsection{The Details-on-Demand Visualization Design Pattern}

Cognitive load has been used to study the effect visual charts can have on users' attention, recognition and working memory, and their impact on judgment and decision-making for critical applications such as weather forecasts \cite{calvo_users_2022} and cockpit information displays \cite{davis_user_2020}. In general, visual content involving heavy use of working memory can negatively impact understanding and, subsequently, the ability to make informed decisions \cite{cairo2012functional}. Some types of data that can be important to visualize but may also add to
cognitive load include uncertainty values \cite{antifakos_evaluating_2004, millet2020hurricane}, temporal variations in values given a snapshot at a particular time \cite{kruiger_multidimensional_2017, wu_graphical_2016, lucjan_perception_2017, hu_motion-based_2025} and  annotations and highlighting over text documents \cite{sun_effect_nodate, yao_effect_2009}. Details-on-demand, a type of interaction design pattern \cite{ahlberg_ivee_1995, schulz_treevis_2011, tableau_data-viz_2016} that shows visual representation of elements upon selection or hovering, can help reduce cognitive load. This method is commonly used in dashboard-style applications, in which people may want a high-level overview of their data but might request additional details about individual points or groups of points by hovering over them (\eg~\cite{yalcin2018keshif}). Common patterns include modal windows, tooltips, and linked views in dashboards \cite{bach2022dashboard}. However, mobile devices and dense displays can pose challenges. 
Another pattern not unique to visualization is animating on hover, such as hovering over an image to activate a GIF \cite{amabili2024show, shu2020makes}, conveying time-varying image data. Adjacent to this, our \st{work looks at}\hl{method} render\hl{s}\st{ing} time-varying numeric values in the form of word-scale line charts \cite{tufte_beautiful_evidence} and similar micro visualizations \cite{isenberg_habilitation}. \st{Deng et al}\hl{Deng et al.} \cite{deng2023revisiting} also describe a similar interaction pattern in their analysis of the design space of composite visualization, terming it an \textit{annotation pattern}.

\subsection{Glanceable Micro-Visualizations}
\label{sec:rw:sparklines}
The representations of quantitative time-oriented data that we reveal in details-on-demand views in this paper recall Tufte's \textit{sparklines} \cite{tufte_beautiful_evidence} and \textit{micro-visualizations} \cite{brandes_micro_visualization,isenberg_habilitation,parnow2015micro}. Brandes \cite{brandes_micro_visualization} and Isenberg \cite{isenberg_habilitation} both define micro-visualizations as ``\textit{very high-resolution visualizations for small- to medium-sized displays}\hl{''}, a context that also applies to our work. Also related is \st{Blascheck et al}\hl{Blascheck et al.}'s \cite{blascheck2018glanceable} definition of \textit{glanceable visualization}, which is associated with a limited availability of time for consumption of visual content, use while in locomotion, having another focus of attention, and smaller display sizes, particularly in the case of smartwatches \cite{blascheck2021characterizing}. In their work on word-scale micro-visualization, Goffin et al \cite{goffin2014exploring, isenberg_habilitation} presented guidelines for their design and placement inline within a body of text. One of these guidelines focuses on dynamic word-scale visualizations which can change in response to interactivity, such as by invoking details-on-demand views. Beck and Weiskopf \cite{beck2017word} similarly characterized different levels of interactivity in word-scale visualization and included details-on-demand as a possible approach, exemplified by a tooltip overlay interaction. Sparklines qualify as glanceable micro-visualization of time-oriented data, although Tufte's original characterization of sparklines \cite{tufte_beautiful_evidence} envisions static word-scale representations that are persistently visible; our use of line charts in this paper encompasses the aforementioned dynamic aspects of micro-visualization revealed in details-on-demand interfaces.

\subsection{Interactive Visualization Beyond the Desktop}

In this paper, we assume multitouch interaction with a touchscreen display.
In the context of visualization on mobile and situated touchscreen displays, there are additional constraints that can serve as sources of cognitive load. For instance, touchscreens often involve direct manipulation of on-screen artifacts \cite{shneiderman_direct_1983}. This naturally leads to some occlusion of visual content, with responsive visualization design tradeoffs \cite{kim2021design} often resulting in the omission of details-on-demand tooltips for small devices. Interaction techniques such as Shift \cite{vogel_shift_2007} have been used to work around this problem, rendering occluded content at an offset to enable users to see it. Similarly, techniques such as 3D Touch and Haptic Touch \cite{noauthor_apples_nodate} offer ways for users to reveal certain contextual details-on-demand using either different levels of pressure exerted on the screen or simply by pressing and holding icons. The other problem associated with touchscreen-based visualization is that of screen size. In \hl{virtual reality (VR)} contexts, screen size is generally correlated with users' cognitive load; for instance, in a training environment, the size of the virtual screen used to render content was found to be correlated with working memory \cite{redlinger_impact_2021}. Participants performed better at memory-based tasks as the screen size increased up to a visual angle of 20\degree, and performed worse as the size became larger. 
A final notable technique specific to communicating the temporal dynamics of data on small displays is the use of looped animation such as GIFs \cite{amabili2024show, shu2020makes}. 
In this paper, we evaluate looped visual animation and temporal haptic feedback for communicating sequences of values in line chart micro-visualizations. 

\subsection{Haptic Feedback \& Visualization}

The sense of touch can also be used to convey information in visualization contexts. A common \hl{use case} for this is in accessibility applications, in which
BLV people can perceive visual information through haptic feedback. For instance, force feedback systems can be used to indicate point positions \cite{braier_haptic_2014, paneels_review_2010, fritz_design_1999}, render haptic glyphs or `hlyphs' \cite{roberts_haptic_2005} or even visualize spatial information such as geographical maps \cite{kaklanis_hapticriamaps_2011}. Related to this, applications such as pin-based shape-changing displays \cite{follmer_inform_2013}, swarm user interfaces \cite{le_goc_zooids_2016} and data physicalization \cite{bastidas_cuya_vibrotactile_2021, herman_touching_2025} all use either active or passive haptic feedback to enhance visually rendered content. Some work has also looked at the learnability of haptic representations of graph structures compared to visual ones \cite{mcgookin2005new}. An important consideration is the use of reinforcing vs. complementary haptic feedback \cite{maclean2017multisensory}; our work focuses on the former, with a brief discussion of how the latter might be implemented.

Haptic perception has also been shown to help people make decisions about spatial reasoning tasks in situations of visual ambiguity \cite{fairhurst2018confidence}, presenting a possible mechanism to support reasoning and decision-making in visual analytics tasks. In the context of immersive analytics, studies have shown that vibrotactile feedback can be used to encourage discussion in collaborative contexts \cite{freeling_vibrotactile_2025} as well as to overcome issues associated with constraints such as the spatial occlusion of points in VR \cite{prouzeau2019scaptics} or supporting guided navigation \cite{han2020exploring, afzaal2025evaluating, alroe_highways_2024}. Work including multimodal emojis \cite{an2022vibemoji} and a wearable device for learning English using phonemes rendered through vibrations \cite{tan_acquisition_2020} have shown that the haptics can be used for affective communication and learning respectively. 
\hl{Structured vibrotactile messages have been introduced in the form of \textit{tactons}, or tactile icons, encoding information through parameters such as rhythm, amplitude, frequency, duration, and location}~\cite{brewster2004tactons,brown2006multidimensional,azadi2014evaluating}\hl{. These parameters can act as visualization channels, although identification degrades as parameter sets become more complex.}
% \mbC{pass-through AR is also constrained by viewing angle}

Beyond perception and affect, we also consider several additional experiential aspects of haptics.
Haptic feedback has been shown to enhance user experience when rendered on top of audio-visual content \cite{maggioni_measuring_2017}, sometimes even in addition to task performance improvements \cite{levesque2011enhancing}. 
Work on the Haptic Experience Inventory (HXI) \cite{kim_defining_2020, shi_development_2025} further standardizes evaluations of haptic feedback in multimedia applications considering both pragmatic and hedonic aspects of experience, akin to Hassenzahl's model of user experience \cite{hassenzahl2007hedonic}. Other forms of haptic interaction have also been shown to promote desirable outcomes, such as enhanced memorability when working with physical visualization \cite{stusak_evaluating_2015}. Thus, there are experiential factors with potential cognitive effects that support the argument for including haptic feedback as an additional information channel for visualization and understanding how user perception, mental models and decision-making are influenced. \st{Drawing from these related bodies of work, we now move the discussion to our design and implementation of vibrotactile line charts.}

Finally, the design space of haptic feedback for visualization is informed by analogous efforts in \st{data} sonification \cite{ludovicosonification, enge2024open}.
An overarching goal of prior sonification research has been to compare alternative techniques \st{so as} to identify suitable perceptual attributes of sound to guide \st{sonification} design  \cite{parseihian2016comparison}. Notably, sonification \st{makes use of}\hl{involves} perceptual attributes \st{that are} also inherent to vibrotactile haptic feedback\st{, namely} \hl{--} frequency and amplitude. \hl{Drawing from these related bodies of work, we now move the discussion to our design and implementation of vibrotactile line charts.}

\section {Design and Implementation}
\label{sec:design}

Our interactions involve touching and holding data points to reveal visual and/or vibrotactile line charts as details-on-demand. Video footage of these interactions can be found in the supplementary material. 
We use scatterplots as a representative \st{visualization} design idiom for our details-on-demand views because these data points can only encode a limited number of attributes in addition to position on the xy-plane. 
The particular combination of a scatterplot with details-on-demand line charts 
has precedents in visual analytics tools such as Keshif \cite{noauthor_comparing_nodate,yalcin2018keshif}, in demonstrations of Orban et al's Drag and Track framework \cite{orban2019drag}, and in work published by \textit{The New York Times}~\cite{ASHKENAS_PARLAPIANO_2014}. 
On tapping and holding a data point in the scatterplot within our interface, we reveal a line chart in a shifted \cite{vogel_shift_2007} tooltip above the touch point. 
When a user lifts their finger off the display, the visualization disappears. This interaction may be implemented in \st{different}\hl{many} ways, e.g., through wearable devices like smartwatches. % However, in the case of wearables an important consideration is the appropriateness of the location because the point of contact with the touchscreen is separated from the point where haptic feedback is felt. 
We render haptic feedback on the index finger touching the display,\st{in line with}\hl{ justified by} Terenti et al.'s findings that users prefer spatially decoupled (\ie distal) haptic feedback on the index finger \cite{terenti_2025_distal}.

\begin{figure*}[h] % [t] = top, can also use [b], [h], [H]
    \centering
    \includegraphics[width=\linewidth]{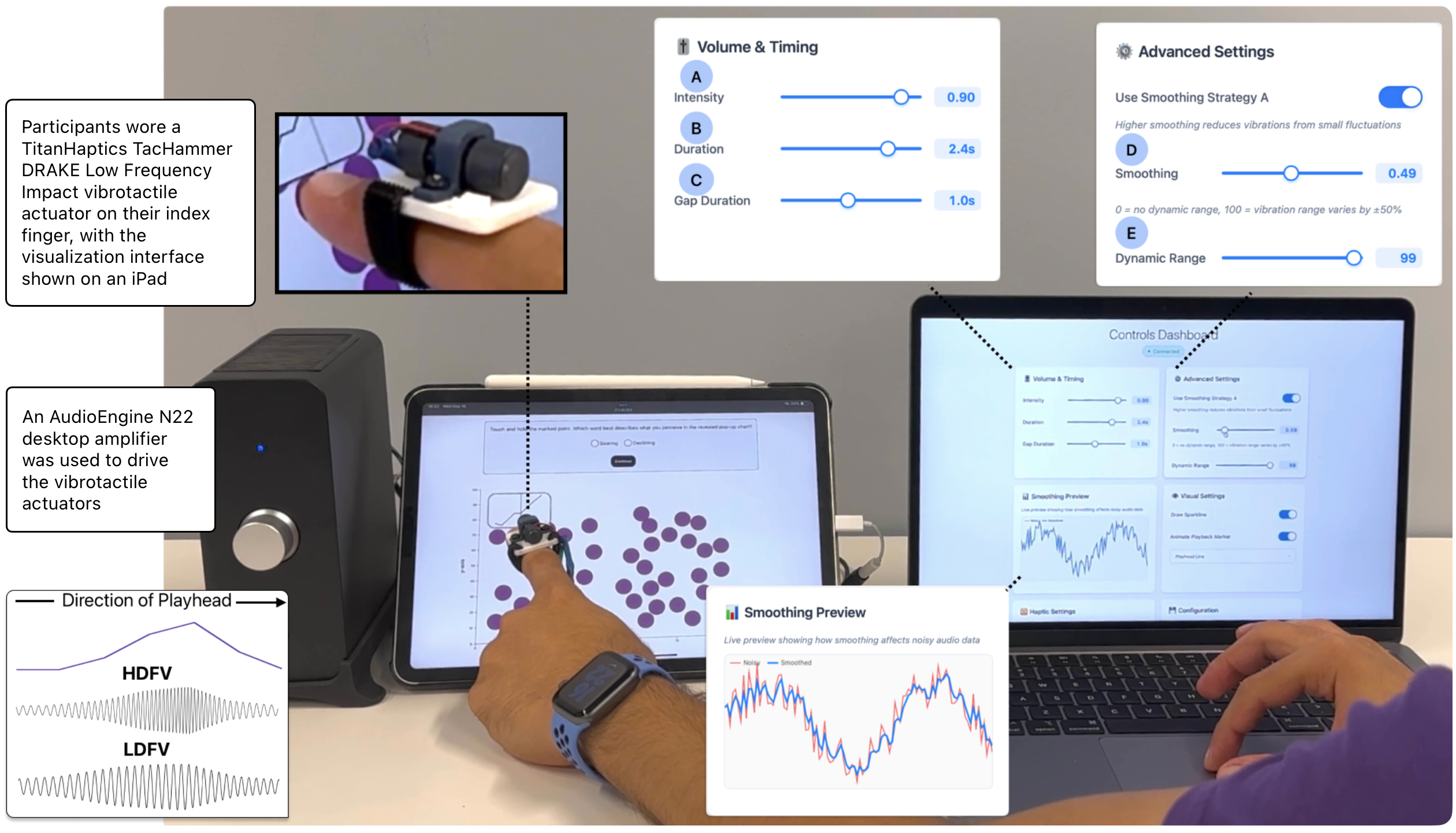}
    \caption{
    The apparatus used for our experiments. The haptic parameter dashboard used in the first phase of Study 1 was accessible using on a laptop. The different parameters exposed in the dashboard are shown above, labelled from A-E. In the bottom left, a diagram of how a line chart is converted to vibrotactile signals. The effects of both low and high dynamic frequency variation on the output frequency are shown. Higher y-values are rendered with greater amplitude and, in the case of high dynamic frequency variation, frequency, as opposed to low dynamic frequency variation, in which amplitude is essentially the only attribute conveying line chart-related information. }
    % \Description[Apparatus]{A photograph of The apparatus used for our experiments. An amplifier (left) was used to drive finger-mounted actuators worn on participants' index finger, with the visualization interface shown on an iPad (centre). The haptic parameter dashboard used in the first phase of study 1 was accessible using on a laptop (right). The different parameters exposed in the dashboard are shown above, labeled from A-E. A represents intensity, B represents duration, C represents gap duration, D represents smoothing, and E represents dynamic range.}
    \label{fig:apparatus}
\end{figure*}

\bstart{Apparatus}
We used the TitanHaptics TacHammer DRAKE Low Frequency Impact vibrotactile actuator \cite{TITANHaptics_2025} for haptic feedback, mounted on a 3D-printed base that was worn on the index finger using velcro straps (see \autoref{fig:apparatus}). The actuator was connected to an Audioengine N22 desktop amplifier, which in turn was connected to an 11-inch iPad Pro. An abstract scatterplot visualization interface, implemented using D3.js and connected to a Node.js server for data logging, was used to reveal line charts on-demand. A 13-inch M3 MacBook Air was used to manipulate haptic settings in an initial algorithm validation study. Participants could tap on a point in the scatterplot to reveal a pop-up chart, which was either purely a visualization, a playhead tracing out haptic playback or a combination of the two. \autoref{fig:teaser} shows how these interactions worked in the context of our visualization interface, while \autoref{fig:apparatus} shows a participant sitting with the experimental apparatus.

% \begin{figure}[h] % [t] = top, can also use [b], [h], [H]
%     \centering
%     \includegraphics[width=0.7\linewidth]{03-interaction-design/figures/ldr-hdr-new-vis.png}
%     \caption{}
%     % \Description[Illustration of visual-to-haptic mapping in our implementation]{A visualization of how a sparkline is converted to vibrotactile signals. The effects of both low and high dynamic frequency variation on the output frequency are shown. Higher y-values are rendered with greater amplitude and, in the case of high dynamic frequency variation, frequency, as opposed to low dynamic frequency variation, in which amplitude is essentially the only attribute conveying sparkline-related information.}
%     \label{fig:ldr-hdr-visualization}
% \end{figure}

\bstart{Haptic rendering}
Haptic rendering \hl{here} involves mapping the y-axis values of \st{a }line chart\hl{s} to an intensity and frequency range that can be used with the haptic actuator, similar to encodings used in prior haptic prototyping applications~\cite{enriquez2003hapticon, swindells2006role}. We determined parameter ranges in pilot studies. \st{We clamped the intensity, derived from the y-values, to a minimum of 25\% to avoid complete `silence' when rendering extremely low values, and normalize within the remaining 75\% of the range. Decibel scaling was used to dictate scaling (supported via the WebAudio API), as is common in audio and haptic applications}~\cite{hollins2003vibratory}. \hl{We imposed a minimum amplitude of 25\% to keep low values perceptible, then interpolated the remaining range in decibel space before passing values to a WebAudio gain node. Because our tasks emphasized pairwise discrimination and trend recognition rather than precise magnitude estimation, equal decibel increments correspond to equal ratios of drive amplitude. This provided a Weber-Fechner-inspired mapping, in line with prior discrimination findings }~\cite{gescheider1990intensity} \hl{and common in vibrotactile and audio applications}~\cite{hollins2003vibratory}\hl{. Additional piloting helped improve comfort and feature saliency. However, we do not claim perceptually uniform scaling for absolute detection, which would require an actuator- and configuration-specific psychophysical magnitude function}~\cite{hwang2013vibrotactile}\hl{ to adhere to a Stevens' Power Law scheme}~\cite{stevens1957psychophysical}.

For TacHammer DRAKE actuators, frequencies between 25--100Hz can be rendered at peak acceleration. \hl{Despite absolute vibration detection thresholds being slightly lower (indicating higher sensitivity) for low-amplitude signals at frequencies between 200-300Hz}~\cite{bolanowski1988four}\hl{, frequency has been found in other work to not affect amplitude discrimination tasks the same way it does absolute detection thresholds}~\cite{gescheider1990intensity}\hl{. Thus, we favored a lower range because it still aligns with rapidly adapting tactile channels while also supporting the actuator's peak acceleration output; higher frequencies were found to cause tingling sensations and discomfort for some pilot participants. We used a base frequency of 50Hz with variations of up to 25Hz, consistent with frequency-discrimination findings}~\cite{mahns2006vibrotactile}. A dynamic frequency variation hyperparameter scaled frequency with y-axis values, following prior visual-to-audio and haptic mappings\cite{yu2000haptic}. We also provided an EWMA/SMA smoothing toggle and coefficient, alongside controls for playback duration, loop interval (250--3000ms), and intensity. \hl{The timing ranges were also informed by prior work looking at vibrotactile signal duration}~\cite{bolanowski1988four}\hl{, temporal-integration}~\cite{gescheider1983temporal}\hl{, and signal gap detection}~\cite{vandoren1990gap}. The bottom-left diagram in \autoref{fig:apparatus} illustrates the resulting signal under high and low dynamic-frequency variation.

% \begin{figure}[h] % [t] = top, can also use [b], [h], [H]
%     \centering
%     \includegraphics[trim={16.5cm 0 0 0}, clip, width=\linewidth]{03-interaction-design/figures/singlehanded-live.png}
%     \caption{An example of single-handed interactions. On pressing and holding a point, its sparkline is revealed as a pop-up and its vibrotactile counterpart is played out, with a visual playhead tracing playback progress.}
%     % \Description[Example of single-point interaction]{Three photographs showing how a participant might interact with our interface using one hand. On pressing and holding a point, its sparkline is revealed as a pop-up and its vibrotactile counterpart is played out, with a visual playhead tracing playback progress.}
%     \label{fig:singlehanded-interaction}
% \end{figure}

% \begin{figure}[h] % [t] = top, can also use [b], [h], [H]
%     \centering
%     \includegraphics[trim={16.5cm 0 0 0}, clip, width=\linewidth]{03-interaction-design/figures/pairwise-live.png}
%     \caption{An example of dual-handed interactions. }
%     % \Description[Example of pairwise comparisons]{Three photographs showing how a participant might interact with our interface using both hands. On pressing and holding a point, its sparkline is revealed as a pop-up and its vibrotactile counterpart is played out on the corresponding finger, with a visual playhead tracing playback progress. When two points are pressed, their vibrotactile playback is synchronized (i.e., in phase).}
%     \label{fig:dualhanded-interaction}
% \end{figure}

\section{Study 1: Vibrotactile Parametrization}

Our first study focused on identifying parameters that could most accurately reflect visual line charts through vibrotactile haptic feedback. We allowed participants to interact with our scatterplot visualization to customize their haptic feedback, after which we evaluated their accuracy for a multiple-choice task to identify which parts of the haptic parameter space impacted task performance. The University of Waterloo research ethics board approved our study (protocol \#47392).

\bstart{Participants}
We recruited 12 participants (5 male, 7 female; all self-reported) for this study, with the inclusion criterion of people above the age of 18 who reported prior experience with data visualizations using software tools such as Excel, PowerBI, Tableau, RStudio or Jupyter Notebook \cite{noauthor_edit_nodate, noauthor_power_nodate, noauthor_tableau_nodate, noauthor_posit_nodate, noauthor_project_nodate}. Participants also had to be able to use touchscreen devices with their index fingers and have normal or corrected-to-normal colour vision. The study took an hour to complete\hl{;}\st{, and} participants were remunerated with a \$20 multi-retailer gift card\st{for their time}. 

\bstart{Dataset}
For the question-answering phase of the task, we used the crowdsourced dataset provided by Bromley and Setlur \cite{bromley_what_2023}. This dataset consisted of 16 unique charts along with 1,892 annotations in the form of descriptor word labels (e.g., rising, falling, peak, valley, etc.) sourced from 67 participants. These descriptors were sometimes assigned to specific parts of the line charts---an x-coordinate was provided in the dataset---but sometimes applied to the global shape of the chart. To design multiple-choice questions around these data points without introducing too much ambiguity, we selected the most frequently occurring label for each chart as the true positive, and randomly selected one of the labels that was never used to annotate a given chart as the true negative, with additional validation using a language model (Claude Sonnet 4) to eliminate words with similar meanings to the true positives. Thus, questions were of the form ``Tap and hold the marked point. Which of the two words best describes that chart that is revealed?'', following a 2-alternative forced choice (2AFC) structure. \autoref{tab:sparklines_summary} shows the list of charts and their corresponding true positives; a complete table with true negatives is included \st{in the}\hl{as} ~\href{https://zenodo.org/records/21953380}{supplemental material}. 

\begin{table}[t]
\scriptsize
\caption{A list of all the unique annotated line charts we obtained from Bromley and Setlur's dataset \cite{bromley_what_2023}, along with their true positive (i.e., the best descriptors for a given chart) labels.}
\newcommand{\tblsize}{\fontsize{6}{7}\selectfont}
\newcommand{\sparkpad}[1]{%
  \makebox[1.25cm][c]{\hspace{-0.3em}\raisebox{-0.5ex}{#1}}%
}
\begin{tabular}{m{0.15cm}|m{1.2cm}|m{1.7cm}|m{0.25cm}|m{1.25cm}|m{1.7cm}}

\textbf{ID} & \textbf{Chart} & \textbf{True Positive Labels} & \textbf{ID} & \textbf{Chart} & \textbf{True Positive Labels} \\
\hline 

\rowcolor{lightgray}
{\tblsize S1} & \sparkpad{\sone} & {\tblsize Bouncing, valley} & {\tblsize S9} & \sparkpad{\snine} & {\tblsize Peak, taking off} \\

{\tblsize S2} & \sparkpad{\stwo} & {\tblsize Gradual, increasing} & {\tblsize S10} & \sparkpad{\sten} & {\tblsize Valley, upturn} \\

\rowcolor{lightgray}
{\tblsize S3} & \sparkpad{\sthree} & {\tblsize Falling, dropping} & {\tblsize S11} & \sparkpad{\seleven} & {\tblsize Tanking, flatline} \\

{\tblsize S4} & \sparkpad{\sfour} & {\tblsize Upturn, taking off} & {\tblsize S12} & \sparkpad{\stwelve} & {\tblsize Bump, slumping} \\

\rowcolor{lightgray}
{\tblsize S5} & \sparkpad{\sfive} & {\tblsize Stable, steady} & {\tblsize S13} & \sparkpad{\sthirteen} & {\tblsize Peak, dropping} \\

{\tblsize S6} & \sparkpad{\ssix} & {\tblsize Peak, bump} & {\tblsize S14} & \sparkpad{\sfourteen} & {\tblsize Plateau, flatline} \\

\rowcolor{lightgray}
{\tblsize S7} & \sparkpad{\sseven} & {\tblsize Flatline, stagnant} & {\tblsize S15} & \sparkpad{\sfifteen} & {\tblsize Decreasing, valley} \\

{\tblsize S8} & \sparkpad{\seight} & {\tblsize Soaring, increasing} & {\tblsize S16} & \sparkpad{\ssixteen} & {\tblsize Tanking, falling} \\

\end{tabular}
\label{tab:sparklines_summary}
\end{table}

\subsection{Procedure}
After obtaining participants' informed consent, the study had two phases, \textsc{free-exploration} and \textsc{question-answering}. 
% \mbC{correct passive voice}

\bstart{Free exploration}
In this phase, we asked participants to interact with the abstract scatterplot interface, revealing different charts generated using the same rule as described by Bromley and Setlur \cite{bromley_what_2023} and feeling the haptic feedback associated with them. At the same time, they interacted with our control dashboard interface, changing the haptic parameter values and evaluating changes in real time, until they arrived at a single configuration that they felt most accurately represented the charts. Each participant's configuration was recorded, along with qualitative responses obtained through a semi-structured interview asking participants about their interactions and personalization strategies.

\bstart{Question answering}
In this phase, we used each participant's preferred configuration, and asked them to answer questions following a 2AFC style. A single point was marked on the scatterplot, and a question asked participants about the best-suited descriptor for the line chart revealed on touching and holding the point. The task was divided into three blocks---(i) \textsc{haptic-only}, (ii) \textsc{visual-only}, and (iii) \textsc{both}, with each block consisting of 10 questions each. Depending on the block, the revealed line chart included (i) only haptic feedback (with a visual playhead indicating which part of the line chart was being rendered through vibrations), (ii) only visual feedback (with a visual playhead but with only a short haptic tick vibration played out), or (iii) both visual and haptic feedback (again, with a visual playhead). We counterbalanced these experimental blocks to account for potential order effects, and recorded participants' task completion accuracy in each of these conditions. Individual line charts did not repeat within a single experimental block, and appeared as a result of uniform random sampling from the set of unused samples.

\subsection{Results}
We conducted a reflexive thematic analysis~\cite{braun2006using} of participant interactions with the parameter customization interface and responses to questions about their parameter space exploration strategies and preferences\st{ for haptic feedback}. Based on our analysis, we synthesize the following themes:

\bstart{T1: Haptic configurations should help people capture maximal information}
Initial exploration strategies looked at individual dimensions one-by-one, with examples such as P5's idea to \textit{``just check what the duration does''}, and \textit{``trying to feel different intensities for the same graph''}. Participants generally gravitated towards strategies that allowed them to gather as much information from the vibrotactile line charts as possible.
Longer vibrotactile line chart playback durations were preferred, with some citing the ability to perceive chart features more clearly (P1, P4, P7, P11). One participant (P1) even expressed their desire for a longer playback duration (5s) than our control interface allowed. On the other hand, two participants (P11, P12) appeared to find a `sweet spot' for duration, with P11 mentioning that duration values above 3s might affect their memory of earlier sections (\textit{``maybe if it is even slower, I'll forget what [played out earlier]''}), even suggesting a possible feature to selectively play back parts of a line chart (\textit{``It would be nice if I can select what part of the line I want to feel''}). While P12 initially felt that longer playback durations were better, they later mentioned that shorter durations were also feasible---in fact, they even perceived shorter playbacks as more intense (final selected duration: 2s). Smoothing values also significantly affected participants' information-gathering (P11: \textit{``I'll leave this at a lower value, because I want to feel all the details''}), with only two participants (P4, P5) selecting smoothing coefficients higher than 0.15 (0.35, 0.55 respectively). These participants also had the worst accuracy (40\%) in the \textsc{haptic-only} condition, hinting at a lack of rendered detail.

\bstart{T2: Some attributes should always be customizable}
From participants' interactions with the parameter customization interface, we saw that most people focused on customizing intensity and playback duration. In the context of intensity, while values were generally close to 1.0, two participants (P1, P12) mentioned that they would likely reduce the intensity over time if the device had to be worn for longer durations (P12: \textit{``I would probably want to do lower intensity if I wanted to use this for a longer period of time''}). This prompted us to always provide a slider control for intensity, consistent with popular haptic tuning mechanisms like D-Box haptic seats \cite{noauthor_d-box_nodate}. On the other hand, while playback duration was assigned a fixed value based on the mean of participants' chosen times, the related attribute of gap duration---controlling the amount of time between looped playback of a line chart---was found to be customized in an unconventional manner. Participants often neglected the parameter (P2: \textit{``I'm just going to ignore gap duration}''),  instead preferring to lift their fingers off the display and press again when they wanted to replay a line chart. Thus, they implicitly controlled the time between successive playbacks for each trial.

\bstart{T3: Frequency variation enhances the expressiveness of higher values, but may reduce it for lower values}
The dynamic frequency factor values showed considerably more variations, with participants initially taking some time to understand what the parameter controlled in the perceived output. Participants found pros and cons with both high and low dynamic frequency variation; while high variation made peaks and other upper-range features `clearer' and easier to perceive (P8: \textit{``I could feel the peak without the visual''}, P10: \textit{``The sharper movements are much easier to detect''}), it negatively affected some participants' perception of low-range values 
% (P9: \textit{``I can tell there's a change in direction and I can tell it's not to the same max, but I could have not told you where it was.''})
(P10: \textit{``The vibrations were definitely harder to feel for lower values.''}). One participant who had a background in computational neuroscience (P7: \textit{``I stop feeling anything after this point, when the vibration is high and then suddenly goes low.''}), hypothesized that this might be because of lower frequencies and intensities combined being harder to perceive right after stronger, higher-frequency vibrations. Moreover, for features that occurred close to the beginning or end of a line chart, some participants had trouble detecting subtle changes in frequency, which made a peak (S9) harder to detect than a similar gradual increase (S2) for instance.

% \begin{table}[h]
% \centering
% \begin{tabular}{ccccc}
% \hline
% Participant ID & Intensity & Duration & Smoothing & Dynamic Frequency Factor \\
% \hline
% 1 & 1.0 & 3.0s & 0.00 & 16 \\
% 2 & 1.0 & 3.0s & 0.00 & 100 \\
% 3 & 1.0 & 2.0s & 0.14 & 38 \\
% 4 & 0.9 & 3.0s & 0.55 & 66 \\
% 5 & 0.9 & 1.5s & 0.35 & 33 \\
% 6 & 0.7 & 1.2s & 0.00 & 0 \\
% 7 & 0.9 & 3.0s & 0.00 & 31 \\
% 8 & 1.0 & 3.0s & 0.00 & 100 \\
% 9 & 0.8 & 1.9s & 0.00 & 19 \\
% 10 & 0.8 & 2.7s & 0.04 & 84 \\
% 11 & 0.6 & 2.2s & 0.00 & 48 \\
% 12 & 1.0 & 2.0s & 0.00 & 20 \\
% \hline
% \end{tabular}
% \caption{\osC{we should complement this with the dashboard or a visual representation of the variables and say which we kept and which we fixed for study 2} \mbC{can this data be represented visually as histograms or dot plots? the table itself feels like supplemental material} The haptic parameter configurations chosen by participants in Study 1. Note that intensity and duration usually took on higher values, and apart from two participants (P4, P5) smoothing values were close to 0. Considerably more variations were seen in the dynamic frequency factor values.}
% \label{tab:haptic_parameters}
% \end{table}

\begin{figure}[b] % [t] = top, can also use [b], [h], [H]
    \centering
    \includegraphics[width=\linewidth]{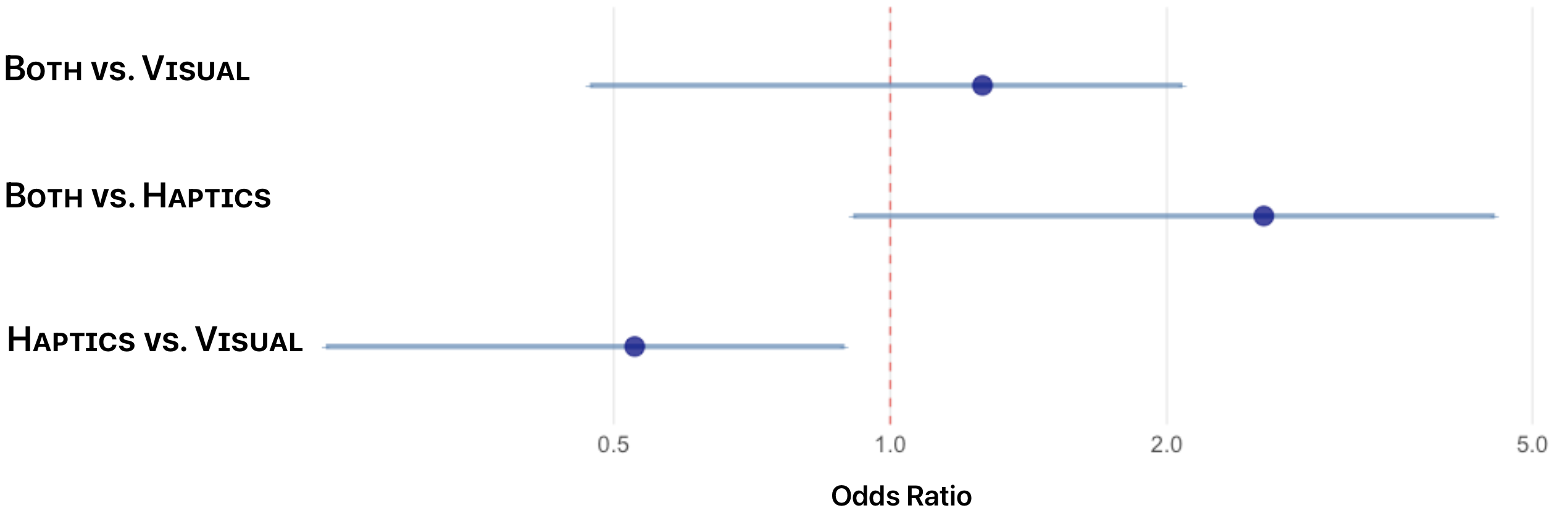}
    \caption{The pairwise log-odds ratios and associated highest posterior density intervals (HPDIs) for Study 1. The \textsc{both} and  conditions had a log-odds ratio close to 1, indicating similar chances of answering correctly in both conditions. On the other hand, participants were twice as likely to answer correctly in the \textsc{both} condition relative to the \textsc{haptic-only} condition (log-odds ratio close to 2), and half as likely to answer correctly in the \textsc{haptic-only} condition as in the \textsc{visual-only} condition. An odds-ratio of 1.0 implies an equal chance of answering correctly in both conditions, while an odds ratio of 0.5 implies that participants are half as likely to answer correctly in the first condition relative to the second.}
    % \osC{explain to the reader what to take away from this chart and/or how to read it; you did this well in fig 7. I remember one reviewer being confused by some of the analyses. Also spell out HPDI as this is the actual first occurrence}}
    % \Description[Pairwise log-odds ratios and associated HPDIs for Study 1]{The pairwise log-odds ratios and associated highest posterior density intervals (HPDIs) for Study 1. The \textsc{both} and  conditions had a log-odds ratio close to 1, indicating similar chances of answering correctly in both conditions. On the other hand, participants were twice as likely to answer correctly in the \textsc{both} condition as compared to the \textsc{haptic-only} condition (log-odds ratio close to 2), and half as likely to answer correctly in the \textsc{haptic-only} condition as in the  condition. An odds-ratio of 1.0 implies participants have an equal chance of answering correctly in both conditions, while an odds ratio of 0.5 implies that participants are half as likely to answer correctly in the first condition as compared to the second.}
    \label{fig:study1-accuracy}
\end{figure}

\begin{figure}[h] % [t] = top, can also use [b], [h], [H]
    \centering
    \includegraphics[width=\linewidth]{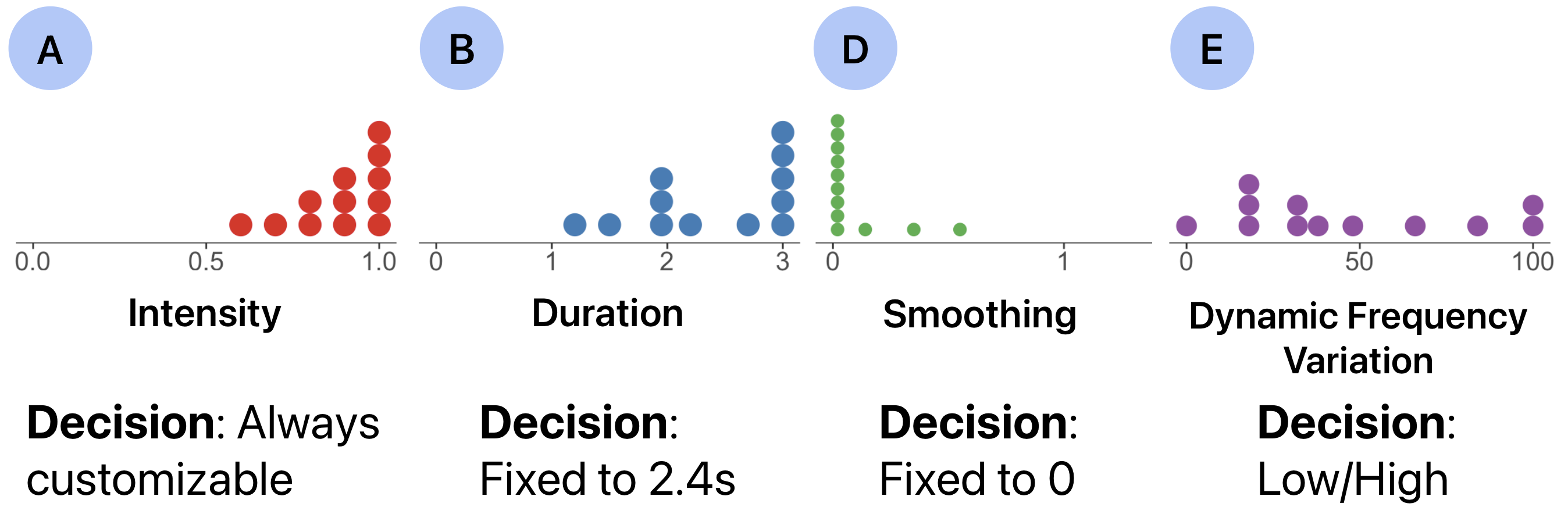}
    \caption{The haptic parameter configurations chosen by participants in Study 1. Note that intensity (A, referencing \autoref{fig:apparatus}) and duration (B) usually took on higher values, and apart from two participants (P4, P5) smoothing values (D) were close to 0. Considerably more variations were seen in the dynamic frequency factor (E) values. Gap duration (C) was ignored by participants and thus was not considered.}
    % \osC{why different colours in the callout histograms It adds noise I think, and further confusion with our dot colours later on as they don't correspond?}}
    % \Description[Haptic parameters]{The haptic parameter configurations chosen by participants in Study 1. Note that intensity (A) and duration (B) usually took on higher values, and apart from two participants (P4, P5) smoothing (D) values were close to 0. Considerably more variations were seen in the dynamic frequency factor (E) values. Gap duration (C) was ignored by participants and thus was not considered.}
    \label{fig:haptic_parameters}
\end{figure}

\bstart{Parametrization results}
We view our quantitative results from two perspectives---the accuracy achieved by participants in each of the three blocks and the parameter values used by them for their haptic feedback configuration. We modeled per-trial accuracy using a Bayesian mixed model with the \texttt{brms} software package, running 4 chains of 4000 iterations for the Markov Chain Monte Carlo algorithm (MCMC). We conducted a prior predictive check; the intercept prior was modeled as \texttt{normal(1.386, 1)}, \st{the }fixed effect coefficients $b$ were modeled as \texttt{normal(0, 0.5)}, and the standard deviation of random effects was modeled as \texttt{student\_t(3, 0, 1)}. Overall, \st{the }posterior estimates were lower for the \textsc{haptic-only} condition (median 84.4\%,  Highest Posterior Density Interval (HPDI) 76.13-91.77\%) than the \textsc{visual-only} (\hl{median} 91.1\%, HPDI 85.87-95.65\%) and \textsc{both} conditions (\hl{median} 92.39\%, HPDI 87.01-96.70\%). \st{The p}\hl{P}airwise log-odds ratios are shown in \autoref{fig:study1-accuracy}.
% conducted Friedman's Test ($p = 0.003$, Kendall's $W = 0.563$ indicating a strong effect size) followed by Dunn's Test with Holm correction, where significant differences were found between participants' accuracy in the \textsc{haptic-only} and  ($p = 0.02$) as well as the \textsc{haptic-only} and \textsc{both} ($p = 0.02$) conditions. 
Looking at individual scores, \st{the }differences are noticeable but not very big, with a drop of 10-20\% in \st{performance }\hl{accuracy} when relying solely on haptics. \autoref{fig:haptic_parameters} shows \st{the }parameter values for each participant. 

We found that participants P4 and P5 were the only ones who selected relatively high values for smoothing, and happened to also be the worst-performing participants in the \textsc{haptic-only} condition. 
As a result, we chose to keep smoothing low for our set of `optimal' configurations, averaging all other participants' values to get 0.02. Playback duration values varied between 1.2 and 3.0 seconds, with no clear correlation with accuracy. We thus used 2.4s, the average of all chosen playback duration values, for our configuration. Lastly, based on dynamic frequency variation values, we created two separate haptic feedback configurations---low and high dynamic frequency variation. These were determined using the average of the lower (12\% frequency variation) and upper (95\% frequency variation) quartiles of dynamic frequency variation values respectively. 
\hl{Study 1 therefore informed the choice to fix smoothing (0.02) and playback duration (2.4s), omitting gap duration as an experimental factor, and carrying low (12\%) and high (95\%) dynamic frequency variation forward as the remaining haptic factor in Study 2.}

\section{Study 2: Simultaneous Comparison}

Next, we evaluated \st{the five haptic feedback configurations obtained from Study 1}\hl{five conditions derived from Study 1}, spanning \visualonly{} (with a short vibrotactile ``tick'' as a control), \haptic{}~\textsc{haptic-only} (two levels of frequency variation: \hapticlow{} and \haptichigh{}) and \both{}~\textsc{both} (two levels of frequency variation: \bothlow{} and \bothhigh{}). This time, participants completed a pairwise comparison task in which they tapped two points at once and answered a question based on the revealed line chart, rendered visually and/or through vibrations. We chose to evaluate simultaneous comparison instead of sequential playback in random order to probe the usefulness of the parallel sensory input our sense of touch affords (i.e., the distance and perceptual separation between index fingertips on opposite hands). Moreover, this helped us avoid any bias due to the rapid decay of haptic memory \cite{shih2009evidence}. We evaluated their performance, along with additional objective and subjective measures. \st{Similar to the first study, t}\hl{T}he University of Waterloo research ethics board approved the study protocol (\#47392).

\bstart{Participants}
We recruited 21 participants (11 male, 9 female, 1 preferred not to answer; all self-reported), all different from those who took part in the first study. Of these, we excluded one participant's data as they repeatedly violated the constraint of simultaneous pairwise comparison. Similar to the previous study, we recruited people who reported prior data visualization software experience. 
We also required participants to be able to use a touchscreen device with both hands' index fingers at the same time, and to have normal or corrected-to-normal colour vision. Sessions took an hour to complete and participants received a \$20 multi-retailer gift card for their time. 

\subsection{Procedure}

After providing informed consent, we asked participants to wear the haptic actuators on both \st{hands'}index fingers, ensuring proper fit and introducing them to the visualization interface. Once they were familiar with the interaction, the question-answering tasks began. There were five blocks in total, with each containing 20 trials. Each trial was of a similar form to Study 1, with a slight difference: instead of choosing between two words for a single point, participants had to choose between two points, one on each side of the display, given a single descriptor word (e.g., \textit{``In the pop-up charts revealed on touching the two marked points, which one is best described by the word `decreasing'?''}). Thus, participants performed a 2AFC task by choosing from two charts. We used the same dataset as Study 1, using 127 distinct pairings of the 16 unique line charts\st{ to use} as trials. This ensured that no trials were repeated \st{even }across conditions in a given session.

The \hapticlow{}, \haptichigh{}, \visualonly{}, \bothlow{}, and \bothhigh{} blocks were presented in counterbalanced order 
and the order of the high and low dynamic frequency variation configurations was counterbalanced 
within the blocks that included haptic feedback. After each block, participants removed the finger-mounted haptic device and responded to the NASA-TLX \cite{hart1988development}, Haptic Experience Inventory (HXI) \cite{shi_development_2025}, and a 7-point semantic differential question asking about confidence in one's responses (\textit{``How confident did you feel in your decisions?''}), in line with prior studies probing confidence \cite{fairhurst2018confidence}. \hl{For each trial, the interface also logged correctness, trial-start and response timestamps (used to derive completion or dwell time), touch events, and completed playback loops.}

\begin{figure}[b] % [t] = top, can also use [b], [h], [H]
    \centering
    \includegraphics[width=\linewidth]{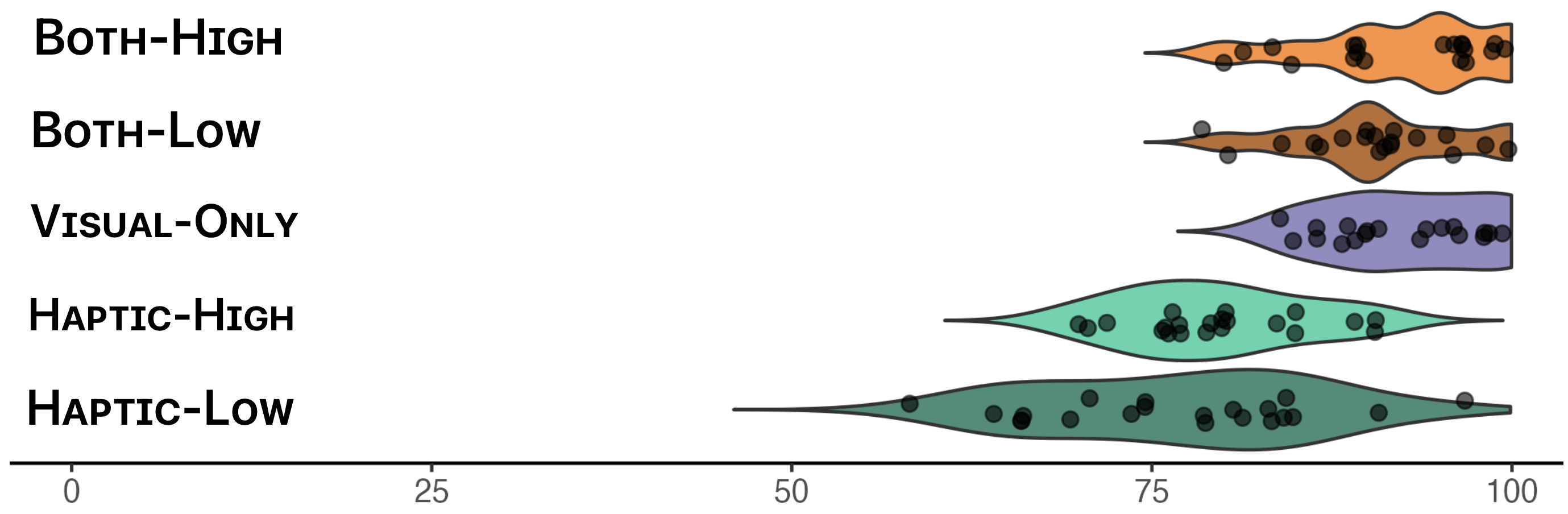}
    \caption{A \st{visualization}violin plot of accuracy scores for Study 2's pairwise comparison task, across each of the 5 conditions.}
    % \Description[Accuracy scores]{A violin plot of accuracy scores for study 2's pairwise comparison task, across each of the 5 conditions. Accuracy was generally lower in the \textsc{haptic-only} conditions than in the  and \textsc{both} conditions.}
    \label{fig:accuracy-plot}
\end{figure}

\begin{figure}[h] % [t] = top, can also use [b], [h], [H]
    \centering
    \includegraphics[width=\linewidth]{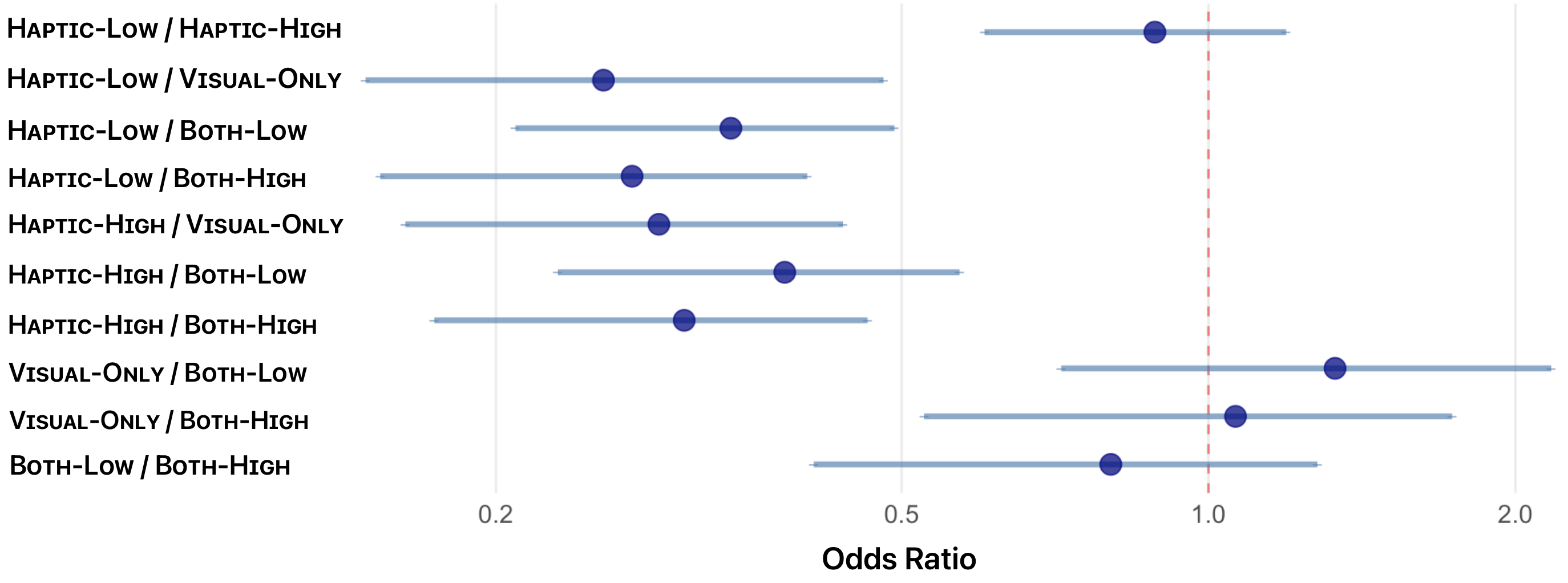}
    \caption{The pairwise log-odds ratios and associated \st{HPDIs} Highest Posterior Density Intervals (HPDIs) for accuracy in Study 2. Overall, odds are worse in the \haptic{}~\textsc{haptic-only} conditions than all other conditions, in which some form of visual feedback is provided. A log-odds ratio value of 1.0 implies that participants were equally likely to answer correctly in both conditions; a ratio of 0.2 means that participants were 20\% as likely to answer correctly in the first condition as compared to the second\st{condition}.}
    % \Description[Pairwise log-odds ratios and HDPIs]{The pairwise log-odds ratios and associated HPDIs for accuracy in Study 2. Overall, odds are worse in the \textsc{haptic-only} conditions than all other conditions, \hl{in which} some form of visual feedback is provided.}
    \label{fig:study2-accuracy}
\end{figure}

% \begin{figure}[h] % [t] = top, can also use [b], [h], [H]
%     \centering
%     \includegraphics[width=\linewidth]{04-study/figures/tlx_violin_plots.png}
%     \caption{A visualization\hl{violin plot} of TLX scores for study 2's pairwise comparison task, across each of the 5 conditions.}
%     \Description[TLX scores for Study 2]{A visualization of TLX scores for study 2's pairwise comparison task, across each of the 5 conditions.}
%     \label{fig:tlx-plot}
% \end{figure}

\begin{figure*}[h] % [t] = top, can also use [b], [h], [H]
    \centering
    \includegraphics[width=\linewidth]{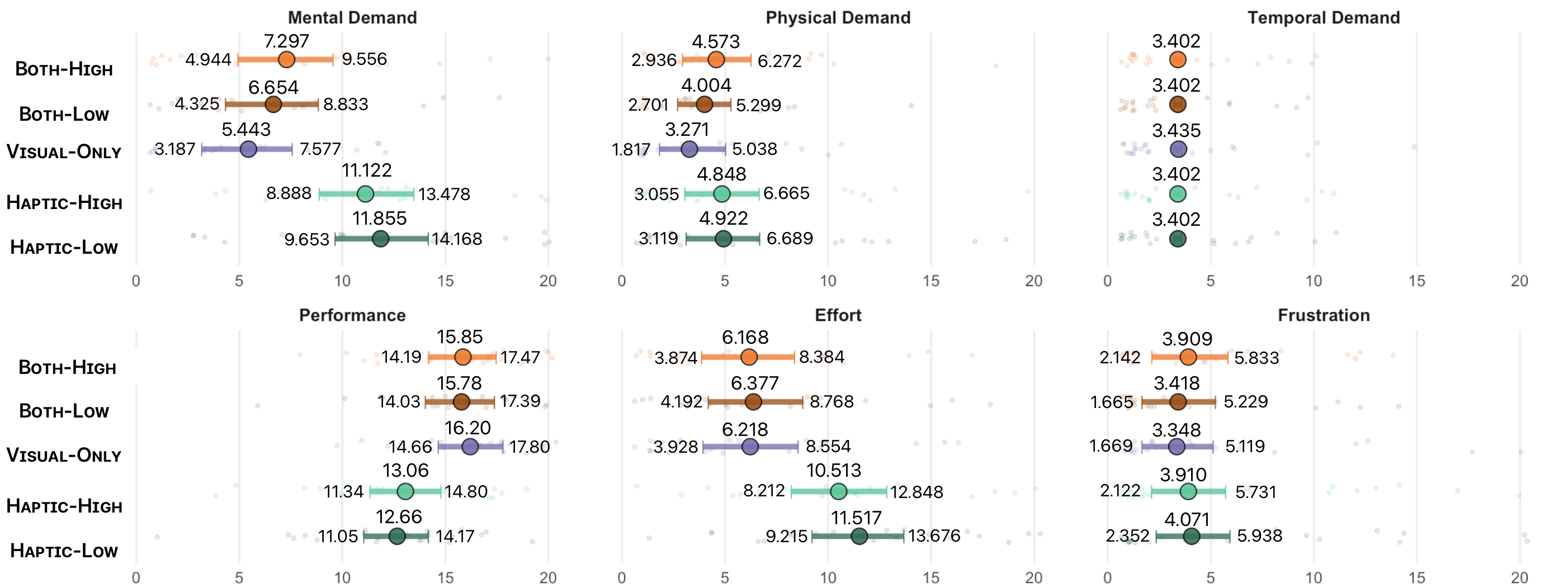}
    \caption{The estimated marginal means and associated HPDIs for TLX responses. Shown as semi-transparent dots in the background are actual participant responses. HPDIs for Temporal Demand are not shown visually due to them being too small.}
    % \Description[Estimated marginal means and HPDIs]{The estimated marginal means and associated HPDIs for TLX responses. Actual participant responses are also shown as semi-transparent dots. We identified non-overlapping HPDIs for mental demand, performance, and effort, between the \textsc{haptic-only} \hapticlow{} \haptichigh{}, \textsc{both} \bothlow{} \bothhigh{}, and  \visualonly{} conditions. Mental demand was higher in the  \hapticlow{} condition compared to the  \visualonly{},  \bothlow{} and  \bothhigh{} conditions; effort saw similar trends between the  \hapticlow{} and  \haptichigh{} conditions compared to the  \visualonly{} ,  \bothlow{} and  \bothhigh{} conditions. For performance, values for  \hapticlow{} and  \haptichigh{} conditions were lower compared to the  \visualonly{},  \bothlow{} and  \bothhigh{} conditions.}
    \label{fig:study2-tlx}
\end{figure*}

\subsection{Results}

We report on both quantitative and qualitative findings, following the `weaving' mixed-methods reporting methodology~\cite{mcchesney2019weaving}. For quantitative analysis, we used the \texttt{brms} package for Bayesian statistics using MCMC with 4 chains of 4000 iterations, similar to Study 1. We compare HPDIs for each model to reach our conclusions, considering measures of accuracy, touch events, task load, haptic experience, and perceived confidence in our analysis. We also conducted an inductive, reflexive thematic analysis~\cite{braun2006using} of participants' descriptions of their experience in the different conditions, revealing three themes that we use to contextualize our results.

\bstart{T1: Haptic perception feels more deliberate, but increases dwell time}
Eight participants stated that haptic feedback created a more grounded experience in combination with visuals (\textit{``The haptic feedback helped by grounding my touch on the screen''}), offering benefits such as being able to describe words and visuals better (P13, P16, P19; \textit{``The word meanings were better communicated through haptics, especially for things like gradual and stable''}), or acting as a double-check on top of visual information (P15, P17, P20). Some participants (P1, P6, P8) reported that these benefits increased over time, as they adapted to the hardware and haptic encoding used. 

\istart{Reinforcing haptic feedback does not change accuracy}
To evaluate how haptic feedback influenced accuracy between conditions, we constructed a logistic regression model for correct answer, using \texttt{normal(0, 1.5)} as the prior for intercept, \texttt{normal(0, 0.5)} for fixed effect coefficients and \texttt{exponential(2)} for random effect standard. \autoref{fig:accuracy-plot} shows the actual accuracy values across each of the five conditions, while \autoref{fig:study2-accuracy}  shows the pairwise log-odds ratios alongside HPDI for each of our conditions. 
The haptics only conditions were lower than the other conditions:
\hapticlow{} (median 76.8\%, 95\% HPDI 72.7--81.2\%) and \haptichigh{} (\hl{median} 79.8\%, 75.3--83.8\%) intervals were lower than the ranges for \bothlow{} (\hl{median} 91.4\%, 88.2--94.1\%), \bothhigh{} (\hl{median} 93.1\%, 90.3--95.7\%), and \visualonly{} (\hl{median} 92.9\%, 89.9--95.4\%).
In all cases, \st{accuracy scores and HDPI are}\hl{accuracy scores and HPDIs were} back-transformed from the logit scale. \st{Because the ranges of both \textsc{haptic-only} overlap, but are separated from the overlapping ranges for all conditions including visual feedback, we conclude that adding haptics on top of visuals was not found to negatively impact task accuracy in a meaningful way.}\hl{The HPDIs for the two \textsc{haptic-only} conditions overlapped each other, as did the three conditions that included visual feedback; the two groups of intervals did not overlap.} Ten participants expressed an \st{interest being}\hl{interest in being} able to play line charts sequentially, instead of performing simultaneous pairwise comparisons.
% (P2-4, P8-10, P12-14,P21)
Despite performing simultaneous comparisons between pairs of points, five participants stated their attention was focused on one vibrotactile line chart at a time. There were no reported issues with the haptic rendering with respect to synchronization with visuals, or input lag. Even when participants had visual feedback in addition to haptics, they tended to let the vibrations play out at least once, sometimes because they felt enjoyable or helped tether the visual content to a tangible sensation (seven participants)
% (P4, P5, P8, P12, P16, P17, P19) 
and other times to actually inform decision-making (seven participants).
% (P1, P2, P3, P7, P13, P15, P20)

\istart{The lack of visual stimuli increases task load, but reinforcing haptic feedback does not} 
We modeled raw NASA-TLX scores as a Gaussian and fit a Bayesian regression model to obtain posterior estimates and pairwise comparisons; we ran 4 chains of 4000 iterations of MCMC. 
We used \texttt{student\_t(3, 0, 30)} and \texttt{student\_t(3, 0, 20)} as the priors for the intercept and $b$ parameters, and \texttt{exponential(1)} for the standard deviation of random effects. 
We identified non-overlapping HPDIs for mental demand, performance, and effort, between the \haptic{} \textsc{haptic-only}, \both{} \textsc{both}, and \visualonly{} conditions. 
Participants reported higher mental demand in the \hapticlow{} 
% \st{(estimated marginal mean 12.01, HPDI 9.88-14.17)} 
condition compared to the \visualonly{} 
% \st{(5.45, HPDI 3.47-7.60)}
, \bothlow{} 
% \st{(6.717, HPDI 4.68-8.98)} 
and \bothhigh{}  
% \st{(7.43, HPDI 5.25-9.67)} 
conditions;
they similarly reported that the \hapticlow{} 
% \st{(11.60, HPDI 9.42-13.73)} 
and \haptichigh{} 
% \st{(10.56, HPDI 8.41-12.85)} 
conditions required more effort relative to the \visualonly{} 
% \st{(6.11, HPDI 3.77-8.26)}
, \bothlow{} 
% \st{(6.68, HPDI 4.38-8.83)} 
and \bothhigh{} 
% \st{(6.38, HPDI 4.21-8.58) }
conditions. 
They also perceived their performance to be poorer in the \hapticlow{}  
% \st{(12.88, HPDI 11.44-14.35)} 
and \haptichigh{} 
% \st{(13.20, HPDI 11.59-14.92)} 
conditions compared to the \visualonly{}  
% \st{(16.45, HPDI 14.92-17.97)}
, \bothlow{} 
% \st{(15.76, HPDI 14.31-17.33)} 
and \bothhigh{} 
% \st{(15.87, HPDI 14.07-17.32)} 
conditions. 
\autoref{fig:study2-tlx} shows the collected TLX scores and estimated marginal means with HPDIs respectively. 
We provide full results including pairwise comparisons in the supplemental material.
% (P2, P3, P8, P9, P14, P18).
Participants' descriptions of the simultaneous perception of two charts varied, with six participants feeling that dissimilar vibrations (\eg soaring vs. falling) were easier to perceive than others.
% (P1, P6, P7, P13, P15, P20)
One participant,
% (P19)
however, had an easier time answering questions when both vibrations were similar with subtle differences instead of completely unlike each other (\textit{``When there is too much going on with the haptics, my brain gets fried''}). 

\istart{Visuals reduce the need for multiple checks, but people still consider haptic feedback}
% To analyze number of touches, 
% \osC{why are we mixing Bayesian and frequentist methods here?} we conducted a Friedman test ($p = 0.0009$, Kendall's W = 0.22) followed by Dunn's test with Holm correction. 
The number of touch events and loops were modeled as a negative binomial variable for bayesian model fitting, with \texttt{normal(0, 2)} and \texttt{normal(0, 1)} \st{being }used as priors for parameters $b$ and Intercept respectively \st{as well as}\hl{, and} \texttt{exponential(1)} for residual and group-level standard deviation. Participants recorded fewer touches for  \visualonly{} (median 2.19, HPDI $1.64 \cdot 10^{-10}$--2.49) than all haptics conditions:  \hapticlow{} (\hl{median} 3.07, HPDI $1.14 \cdot 10^{-8}$--3.47),  \haptichigh{} (\hl{median} 3.14, HPDI $5.54\cdot10^{-10} - 3.55$),  \bothlow{} (\hl{median} 2.73, HPDI $1.82\cdot10^{-12}$--3.10) and  \bothhigh{} (\hl{median} 2.82, HPDI $6.26\cdot10^{-10}$--3.20). Similarly, participants replayed vibrotactile line charts fewer times in the  \visualonly{} (\hl{median} 0.94, HPDI 0.73--1.20) condition (in which haptic feedback was being limited to a single tick being played per tap), compared to \st{the }\hapticlow{} (\hl{median} 5.43, HPDI 4.23--6.75), \haptichigh{}  (\hl{median} 5.79, HPDI 4.54--7.22), \bothlow{} (\hl{median} 3.28, HPDI 2.50--4.01) and \bothhigh{} (\hl{median} 3.60, HPDI 2.81--4.50) conditions. %:  and  participants recorded significantly more plays than  ($p < 0.0001$ for both comparisons). The  and  conditions also recorded significantly more plays that 
% ($p < 0.0001$ for all comparisons). \autoref{fig:todo:touch event}

\st{We do not report on raw time taken because external factors such as readjusting the finger mount and changing the intensity of the actuators added noise to time measurements.}\hl{Because readjusting the finger mount and changing actuator intensity added noise to raw completion times, we did not use absolute time to compare conditions. However, we retained timing data as an exploratory dwell-time measure: we saw that both participant-level median dwell time and HXI involvement scores increased in each condition containing vibrotactile feedback, but not in} \visualonly{}.

\bstart{T2: Vibrotactile line charts inform mental models of words and visuals}
Echoing points raised by participants in Study 1, participants reported trying to trace out or draw mental images of line charts based on the vibrations they felt. This occurred for both participants who started off with one of the conditions with visual information (\ie~\visualonly{}, \bothlow{}, and \bothhigh{}) (14 participants)
% (P1, P2, P3, P4, P7, P9, P10, P12, P13, P14, P17, P18, P19, P21)
as well as those who started with \hapticlow{} or \haptichigh{} (7 participants). In \hapticlow{}, \haptichigh{} conditions, participants typically dwelled for longer, with 12 people reporting that they spent time trying to create mental images of line charts from the verbal description mentioned in the question and the vibrations they felt. P13 reported feeling that word meanings were \textit{``better communicated''} when haptic feedback was present.

With visual feedback included, P10, P11 and P12 mentioned that for certain visuals, the vibrations made them second-guess themselves due to ``feeling different'' from their initial expectations. P11 also mentioned that they relied on intuition much more when they only had access to one modality as opposed to slower thinking when both visuals and haptic feedback were available. 
% (P5, P6, P8, P11, P15, P16, P20). 
Those who had already seen completed blocks of trials with visual information reported being able to match vibrations to features that they had already seen and felt in prior cases (P1, P7, P12). As seen in Study 1, the dynamic frequency variation factor played a significant role in shaping the mental image created by vibrotactile line charts---while features typically found in the upper end such as peaks and plateaus were perceived to be `clearer' and `sharper' (P13, P19) with high dynamic frequency variation, lower end values were harder to feel (P10, P18). On the flipside, some participants felt that low dynamic frequency variation offered more benefits in the form of enhanced low-range clarity than disadvantages such as the inability to detect cases in which intensity only varied slightly, which P13 described as drawing a \textit{`softer'} sketch of the line chart in their mind. Also worth noting is that although mental images typically helped participants make decisions, in ambiguous cases they also added cognitive load because participants either felt that both vibrations were too similar to tell apart, or did not perceive the vibration to faithfully match the visual line chart. 
It is important to note that the presence of visual feedback was also important, as participants would otherwise often either close their eyes, look up at the ceiling or otherwise look away from the display when only haptic feedback was rendered, despite the presence of a visual playhead. In fact, without tethering vibrations to visuals, P19 reported \textit{``I was more prone to getting distracted by the vibrations on the other finger''} or \textit{``losing track of time''}, indicating that the amount of interference between the perception of vibrations on both fingers depended on the presence of visual stimuli, and that even the playhead's position helped them align \st{the mental image of a}line chart\hl{s} with vibrations.

\begin{figure}[h] % [t] = top, can also use [b], [h], [H]
    \centering
    \includegraphics[width=\linewidth]{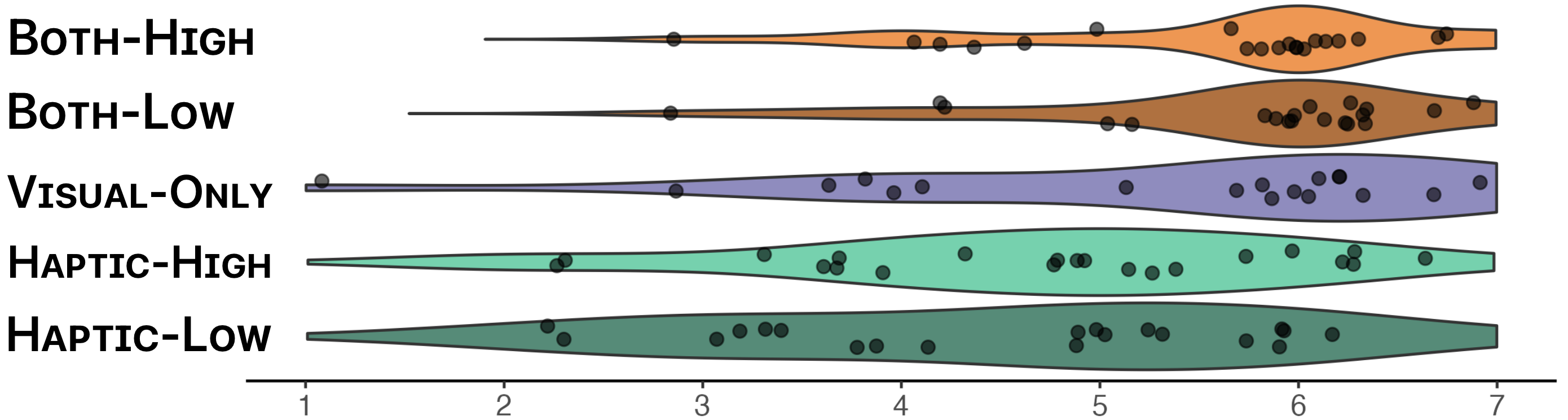}
    \caption{A violin plot of the confidence values (ranged 1-7) reported by participants for each of our 5 conditions.}
    % \Description[Confidence values]{The confidence values (ranged 1-7) reported by participants for each of our 5 conditions. Participants were generally less confident in the \textsc{haptic-only} conditions than in the others, although a wide spread was seen.}
    \label{fig:confidence-plot}
\end{figure}

\istart{Vibrotactile line charts do not affect confidence in responses}
\autoref{fig:confidence-plot} shows the confidence values participants responded with for each of our 5 conditions. Analyzing confidence scores \st{using}a cumulative model, we used \texttt{student\_t}(3, 0, 1.5) and \texttt{student\_t}(3, 0, 5) as priors for the $b$ and intercept parameters respectively, and \texttt{exponential}(1) for random-effect standard deviation. We saw \st{that }expected scores for \hapticlow{} (median expectation 4.76, HPDI 4.12--5.36) and \haptichigh{} (\hl{median} 4.64, HPDI 3.96--5.26) were lower than the \st{conditions with  }\visualonly{} (\hl{median} 5.61, HPDI 4.96--6.09), \bothlow{} (\hl{median} 5.72, HPDI 5.22--6.10) and  \bothhigh{} (\hl{median} 5.64, HPDI 5.12--6.05)\hl{ conditions}; however, the \st{overlap between }\hl{overlapping} HPDIs suggest\st{s} \st{that adding}reinforcing \st{haptic feedback}\hl{haptics} does not lower confidence in participant decisions. 

\begin{figure*}[h] % [t] = top, can also use [b], [h], [H]
    \centering
    \includegraphics[width=\linewidth]{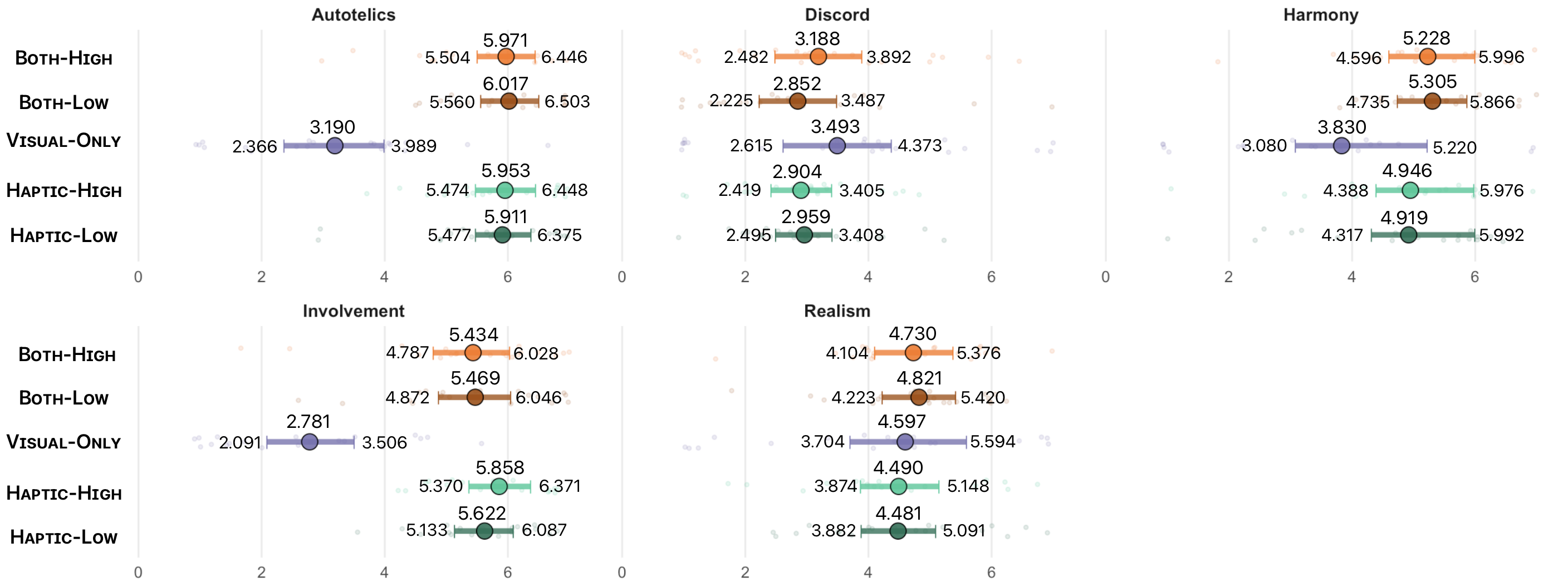}
    \caption{The estimated marginal means with HPDIs for each dimension of the HXI based on responses collected in Study 2, across each of the 5 conditions. Also shown as semi-transparent dots in the background are actual participant responses.}
    % \Description[Estimated marginal means and HPDIs]{The estimated marginal means with HPDIs for each dimension of the HXI based on responses collected in Study 2, across each of the 5 conditions. The  condition has lower marginal mean values for Autotelics, Involvement and Harmony.}
    \label{fig:study2-hxi}
\end{figure*}

\bstart{T3: Dynamic frequency variation as an experiential dimension}
% \osC{awkward and wordy intro} 
Looking at the role of dynamic frequency variation, we found \st{that}certain features were easier to perceive with high dynamic frequency variation than others. This included features such as `peak', `valley', `bouncing', and `gradual'; participants felt that \st{the}sharp extreme points were much easier to detect when both intensity and frequency varied. Even in the case of `valley' and `gradual', in which a significant portion of the line chart was in the low-range, the transition between the upper and lower ranges made the vibrotactile renderings easy to identify. While the bottom-most point of a valley may have felt like complete silence, it being \st{surrounded}preceded by a sharp drop and immediately followed by a sharp increase accentuated the minimum. Similarly, the fact that the `gradual' line chart began with the low range and then linearly increased in magnitude made participants perceive a gradual increase in intensity and frequency (P6). In terms of accuracy, we see that high dynamic frequency variation might yield slightly better accuracy (1.7--2.1\% for median), but this is low compared to uncertainty in the estimates.  Other participants reported noticeable improvements for these words when using low dynamic frequency variation, claiming that while \st{the}features felt `duller' or `smoother' than with high dynamic frequency variation, the signal was much easier to perceive. In terms of perceived confidence, however, we did not observe lower scores for low dynamic frequency variation in \hapticlow{} or \bothlow{} conditions compared to \haptichigh{}, \bothhigh{}, or \visualonly{} conditions.

\istart{Vibrotactile line charts promote higher involvement}
To further probe the experiential contributions of haptic feedback, we modeled HXI scores similar to TLX scores, fitting a Bayesian regression model to obtain posterior estimates and pairwise comparisons; once again we ran 4 chains of 4000 iterations of MCMC. We used \texttt{student\_t(3, 0, 1.5)} and \texttt{student\_t(3, 0, 1.0)} as priors for the intercept and $b$ parameters, and \texttt{exponential(1)} for the standard deviation of random effects. Looking at posterior estimates, we found that the Autotelics dimension was rated higher in \hapticlow{}, \st{(estimated marginal mean 5.806, HPDI 5.33-6.26)}  \haptichigh{}, \st{(5.84, HPDI 5.35-6.31)} \bothlow{}, \st{(5.97, HPDI 5.54-6.42)} and  \bothhigh{} \st{(5.90, HPDI 5.42-6.34)} conditions compared to  \visualonly{}. \st{(3.17, HPDI 2.36-3.92)}
Similarly, the inclusion of haptic feedback resulted in higher Involvement scores \st{ \hapticlow{} }\st{(Involvement: 5.64, HPDI 5.19-6.10; Harmony: 4.95, HPDI 4.36-5.50),}\st{ \haptichigh{} }\st{(Involvement: 5.87, HPDI 5.41-6.35; Harmony: 4.88, HPDI 4.27-5.50),}\st{ \bothlow{} }\st{(Involvement: 5.54, HPDI 4.98-6.14; Harmony: 5.35, HPDI 4.78-5.89)}\st{and  \bothhigh{} }\st{(Involvement: 5.46, HPDI 4.83-6.04; Harmony: 5.25, HPDI 4.64-5.85)}compared to  \visualonly{}. \st{(Involvement; Harmony: 3.72, HPDI 2.87-4.55)}
Conditions with reinforcing haptic feedback; i.e.,  \bothlow{} and  \bothhigh{} received higher Harmony ratings than  \visualonly{}. \autoref{fig:study2-hxi} depicts estimated marginal means and corresponding HPDIs, as well as participant responses. All HXI dimensions had a \st{Crohnbach's}\hl{Cronbach's} Alpha greater than 0.88, indicating high reliability. Detailed pairwise comparison results are included as supplementary material.

\section{Discussion}

We now reflect further on our study findings, offer implications for adding vibrotactile feedback to visualization interfaces, and comment on the limitations of our work.

 \bstart{Adding vibrotactile feedback to line charts shows experiential benefits without reducing task accuracy}
In both studies, we observed that participants performed similarly in the \visualonly{},  \bothlow{}, and \bothhigh{} conditions; they were less accurate in \hapticlow{} and \haptichigh{} conditions, in line with prior work \cite{lobb1965vision, fairhurst2018confidence}.  
\hl{In Study 2, the overlapping accuracy intervals for} \visualonly{}, \bothlow{}, and \bothhigh{} \hl{support the interpretation that adding reinforcing haptics did not meaningfully reduce accuracy relative to visual feedback alone.}
As reflected in our thematic analysis and participants' HXI responses, there was also a distinct experiential benefit associated with vibrotactile feedback, corresponding specifically to higher involvement when haptic feedback was present, similar to prior investigations that added vibrotactile feedback to audio-visual experiences \cite{maggioni_measuring_2017} and emojis \cite{an2022vibemoji}. Such experiential benefits are also thought to improve memorability, echoed in prior work on data physicalization~\cite{stusak_evaluating_2015} and vibrotactile vocabularies~\cite{tan_acquisition_2020}\hl{; probing the effectiveness of our vibrotactile line charts for improving learning outcomes offers a direction for further investigation}.

\bstart{We hypothesize that vibrotactile feedback could elicit more deliberate decision-making}
Adding vibrotactile feedback resulted in similar levels of perceived confidence and cognitive load relative to a \visualonly{} condition. 
\st{However, it was also associated with slower response times, attributed by some participants to the hedonic dimensions of haptic experience where participants would let vibrations play out multiple times just because they felt pleasant.}\hl{Our exploratory dwell-time analysis associated longer dwell with higher involvement in every condition containing line-chart haptics; participants also replayed more loops and described sometimes letting pleasant vibrations finish before answering.} Thus, additional attention was attributed to the hedonic aspects of the vibrations, as reflected in the higher autotelics scores in all conditions relative to \visualonly{}. In some other cases, participants reported experiencing discrepancies in visual and haptic perception.
These discrepancies forced them to dwell on the visual and vibrotactile stimuli for longer, prompting them to reconsider their initial response. 
This is in agreement with prior work on combined visual and haptic perception, in which it was found that haptic sensations influence the perception of visual imagery~\cite{blake2004neural}. 
This was observed to be true in cases in which both egocentric haptic cues and environmental cues are provided, or haptic experience is otherwise more grounded and salient with the visual representation \cite{kelly2011haptic}.
We also found that increased task completion time was associated with higher reported involvement for all conditions other than \visualonly{}.  
This additional quanta of attention also appeared to depend on the specific type of comparison being made; for instance, detecting a steep rise or fall was often perceived to be easier than detecting a stable or gradual trend. 
The potential ambiguity of the feature descriptors we employed~\cite{bromley_what_2023} also likely influenced response deliberation. 
Overall, this deliberation might be viewed from the lens Kahneman refers to as System 1 and System 2 thinking~\cite{kahneman2011thinking}, \st{in which}\hl{where} haptic feedback can trigger a switch from \st{what was}\hl{a task} typically described by participants as \st{a}`quick' and `simple' (i.e., System 1) \st{task}in the presence of visuals alone to a System 2 task involving more deliberate thinking with the addition of vibrotactile feedback, possibly due to the way \st{haptic feedback}\hl{vibrations} influenced \st{participants’ perception}\hl{the perception} of \st{the line} chart\hl{s} and \st{its}\hl{their} temporal features.

\subsection{Future Work and Implications for Design}

\bstart{`Slow analytics' experiences may present a viable scenario for reinforcing vibrotactile encoding}
\st{Whether due to hedonic experience or perceived discrepencies, the aforementioned deliberation seems to be particularly well-suited for applications designed to support `slow analytics' use cases}\hl{The dwell-time pattern, combined with the increase in involvement in the presence of haptic feedback, whether due to hedonic experience or perceived dissimilarities in sensation, appears to suggest that reinforcing haptics can support `slow analytics'}~\cite{bradley2019approaching}: a close reading of data as described by digital humanities scholars. 
Interfaces that allow people to both see and feel individual data points similarly suggest \st{an}alignment with data humanist values \cite{lupi2017vis}, such as when \st{every }data point\hl{s}\st{ one touches } represent\st{s} a human individual.  

\renewcommand{\sparklineheight}{2.25}

\bstart{\st{Adding haptic icons to line charts could draw further attention to important temporal features}\hl{Overlaying tactons could draw attention to important temporal features}}
\label{subsec:haptic-icons}
We noted that dynamic frequency variation played an important role in shaping participants' experiences with the interface, with some participants describing perceivable differences in the mental images generated by vibrotactile line charts based on frequency variation. 
However, we found no major differences in accuracy or haptic experience when comparing the two approaches. 
For some, the benefits afforded by high dynamic frequency variation \st{including}\hl{included} improved clarity for features occurring at higher quantitative values. 
In contrast, other participants noticed a consequent loss of clarity in the lower range of data values. 
Adapting the frequency variation based on the current data range could improve clarity at both ends, for example, by using a nonlinear mapping that varies frequency for high values but keeps it static for lower data values.
\st{Alternatively, a simpler solution (and one mentioned by P7 in Study 1) may involve overlaying a line chart with haptic `icons': short, distinct vibrotactile sequences to indicate features such as global maxima or minima.}\hl{P7's proposed short sequences for extrema are related to the concept of \textit{tactons}}~\cite{brewster2004tactons}\hl{. They would add a categorical channel for extrema or threshold crossings atop the continuous amplitude and frequency channels.}
This recalls the visual icons sometimes used in Tufte's sparklines~\cite{tufte_beautiful_evidence}, such as blue and red markers for extreme values: \sonedotted. 
\hl{Like VibEmoji}~\cite{an2022vibemoji}\hl{, this pairs parameterized vibration with visual symbols; here, sparse tactons annotate a continuous data representation. Because multidimensional tacton recognition declines as parameter values multiply}~\cite{brown2006multidimensional}\hl{, designers should use small, empirically separable vocabularies}~\cite{qian2009distinguishable}\hl{ and prioritize temporal pattern over additional amplitude-like dimensions}~\cite{azadi2014evaluating}\hl{, potentially even parameterizing icons based on data attributes}~\cite{israr2014feel}\hl{.}
The overlay approach also aligns with participants' tendency to extract the most granular, detailed information from what was available to them through the rendering algorithm. 
As with the personalization preferences we observed in Study 1, people may be inclined to customize \st{their own haptic icons}\hl{tactons} based on their own perceptual thresholds and preferences~\cite{seifi2017exploiting}. 

\bstart{Vibrotactile scented widgets could direct attention in subsequent visual analysis}
Participants' task performance was worse when visual stimuli were absent, with a greater dip in Study 2 (with accuracy range 60--95\% as opposed to 70--100\%). 
This points towards an increased difficulty in perceiving the charts through touch alone, at least for sighted people. 
However, this difference may not be practically significant, particularly for single-point interactions, such as instances in which participants only inspected a single data element in the chart. 
This suggests that vibrotactile encoding can serve as `information scent' \cite{pirolli1998information,willett_scented_2007} to direct attention toward trends not shown in full visual detail. 

\st{Vibrotactile encoding could mitigate visual ambiguity}\bstart{\hl{Potential directions for further evaluating the perception of vibrotactile encodings}}
When visualizing volumetric spatial data in immersive analytics contexts
\cite{ens_grand_2021}
, reinforcing vibrotactile encoding \hl{can circumvent} \st{not be affected by} \hl{the issue of} viewing angle distortion
~\cite{zhan2020augmented, tong2020optical}
. Future studies could look at vibrotactile encoding for redundantly representing attributes mapped visually via position and colour channels, as both of these channels are susceptible to distortion. The viability and relative effectiveness of haptic feedback should also be evaluated and compared to audio, as prior work on `eyes-free use' \cite{oakley2007designing, li2008blindsight} has explored sonification-based approaches.   Looking at participants' ability to simultaneously perceive two vibrotactile signals also poses the question of whether a haptic analogue exists for the `\textit{eyes beat memory}' visualization design principle~\cite{munzner2014visualization}: if juxtaposing multiple views of data is generally preferred over animating or paginating between states, we should determine if the perception of haptic data encoding can be similarly parallelized. 
Haptic memory is known to be extremely limited and similar in duration and decay to iconic visual memory \cite{shih2009evidence}. 
However, unlike sonification and auditory stimuli, haptic feedback can be rendered on different parts of the body, which potentially alleviates perceptual masking. \hl{The effect of reinforcing vibrotactile representation on memorability presents a similar direction for future work.}
% ~\cite{ens_grand_2021,zhan2020augmented,tong2020optical}.

\st{Vibrotactile encoding could represent secondary attributes}
\st{Given the relatively poorer task performance associated with the haptic-only conditions, we hesitate to suggest the vibrotactile encoding of data attributes that are complementary to those shown visually.}
\st{However, we remain optimistic about the vibrotactile encoding of secondary or derived values that qualify those shown visually.}
\st{For instance, the degree of uncertainty around predicted time-series values could be felt rather than seen, as an alternative to showing visual uncertainty around point estimates using coloured bands or error bars.}

\st{Additional secondary values to consider for vibrotactile encoding manifest in collaborative analysis scenarios}
% ~\cite{mahyar2014supporting}.
\st{While real-time collaborator activity such as their current area of interest or their selected data points could be represented with additional visual indicators or annotations on selected marks, this information could also be felt via vibrotactile encoding.}

% to communicate quantitative time-series data as in our use-case, or even to communicate remote collaborator activity in a shared virtual workspace. 

\bstart{Applications suggested by our experimental interface}
% Based on our empirical findings, we know that (1) visualizations augmented with redundant haptic feedback draw users' attention more and lead to increased dwell time, often helping ground what they see visually or even adding context to verbal descriptions of charts, and that (2) haptic feedback alone can also be used to communicate time-varying values, albeit at a reduced perceptual resolution. 
While the stimuli shown in our interface were motivated by analytical tools
~\cite{orban2019drag,yalcin2018keshif}
and instances of data storytelling~\cite{ASHKENAS_PARLAPIANO_2014}, we anticipate based on prototypes we developed that similar details-on-demand interfaces could be developed around electoral polling maps, where trends for jurisdictions could be revealed in tooltips.
Another interface, and one particularly familiar on mobile devices, is a list of stock price values and trend indicators, where line charts of price fluctuations and trading volumes are revealed on demand.

\subsection{Limitations} 
Some limitations of our work include generalizability to populations with lower visualization literacy, the inability to freely scrub line charts in either temporal direction, and the need for a wired connection to the haptic actuators to ensure low latency. 
\section {Conclusion}
We investigated vibrotactile feedback as a reinforcing information channel in \textit{details-on-demand} views. We specifically used this feedback to encode continuous time-oriented data rendered visually as glanceable line chart micro-visualizations, informed by a study in which participants explored the vibrotactile parameter space to obtain configurations that they perceived to most accurately match visual stimuli. With five distinct haptic feedback configurations, we then conducted a second study\st{in which}\hl{.}\st{w} \hl{In this study, w}e evaluated participants' performance on a pairwise comparison task as well as their perceived cognitive load, haptic experience, and confidence in question-answering. Adding vibrotactile feedback to details-on-demand views \hl{was found to} result in similar levels of task performance\hl{.} \st{with p}\hl{Notably, p}articipants remark\st{ing}\hl{ed} that \st{it}\hl{the reinforcing vibrotactile feedback} captured their overt attention, which we hypothesize to be a precursor to more deliberate decision-making.
% Based on our findings, we discussed three representative use-cases that demonstrate the addition of vibrotactile feedback to visualization applications, either in details-on-demand views or to complement glanceable summaries on small displays. 

%% if specified like this the section will be omitted in review mode
% \acknowledgments{%
% 	The authors wish to thank A, B, and C.
%   This work was supported in part by a grant from XYZ (\# 12345-67890).%
% }

% \bibliographystyle{abbrv-doi-hyperref}
\bibliographystyle{abbrv-doi-hyperref-narrow}
\setlength{\bibspacing}{-0.35pt}
\bibliography{DataVibes_shortdoi_abbrev}

\end{document}